\documentclass[twocolumn,pra,article,showpacs,superscriptaddress]{revtex4-1}
\usepackage{bm}
\usepackage{epsfig}
\usepackage{graphicx}
\usepackage{amssymb}
\usepackage{amsthm}
\usepackage{cancel}
\usepackage{amsmath}
\usepackage{color}
\newcommand{\comments}[1]{}

\newcommand{\omegar}{\omega^{\textrm{r}}}
\newcommand{\omegai}{\omega^{\textrm{i}}}
\newcommand{\mueff}{\mu}
\newcommand{\neff}{n}
\newcommand{\F}{\hat{\mathbf{F}}}
\newcommand{\Fx}{\hat{F}_x}
\newcommand{\Fy}{\hat{F}_y}
\newcommand{\Fz}{\hat{F}_z}
\newcommand{\I}{\hat{\mathbb{I}}}
\newcommand{\HH}{\hat{H}}

\newcommand{\U}{\hat{U}}
\newcommand{\Btwo}{\hat{B}_2}
\newcommand{\Bfour}{\hat{B}_4}
\newcommand{\h}{h}
\newcommand{\absh}{|h|}
\newcommand{\deltah}{\delta_{\h}}
\newcommand{\cc}{r}
\newcommand{\absc}{|r|}
\newcommand{\deltac}{\delta_{\cc}}
\newcommand{\kp}{k_{+}}
\newcommand{\km}{k_{-}}
\newcommand{\Fzexp}{\langle\hat{F}_z\rangle}
\newcommand{\thetaMag}{\theta}
\newcommand{\thetaSpin}{\tilde{\theta}}
\begin{document}

\title{Excitation spectrum of a toroidal spin-$1$ Bose-Einstein condensate}

\author{H. M\"akel\"a}
\affiliation{QCD Labs, COMP Centre of Excellence, Department of Applied Physics, Aalto University,
P.O. Box 13500, FI-000 76 AALTO, Finland}
\affiliation{Department of Physics, Ume\aa \,\,University, SE-901 87 Ume\aa, Sweden}
\author{E. Lundh}
\affiliation{Department of Physics, Ume\aa \,\,University, SE-901 87 Ume\aa, Sweden}

\begin{abstract}
We calculate analytically the Bogoliubov excitation spectrum of a toroidal spin-1 Bose-Einstein condensate that is  subjected to a homogeneous magnetic field and contains vortices with arbitrary winding numbers in the $m_F=\pm 1$ components of the hyperfine spin. 
We show that a rotonlike spectrum can be obtained, or an initially stable condensate can be made unstable by adjusting the magnitude of the magnetic field or the trapping frequencies. The structure of the instabilities can be analyzed by measuring the particle densities of the spin components. We confirm the validity of the analytical calculations by numerical simulations. 
\end{abstract}


\maketitle
\section{Introduction}\label{sec_Int}
Bose-Einstein condensates (BECs) confined in toroidal 
traps have been the subject of many experimental studies recently~\cite{Ryu07,Ramanathan11,Moulder12,Wright12,Marti12,Beattie13}.
This research covers topics such as the observation of persistent current~\cite{Ryu07}, phase slips across a stationary barrier~\cite{Ramanathan11}, stochastic~\cite{Moulder12} and deterministic~\cite{Wright12} phase slips between vortex states, the use of toroidal condensates in interferometry~\cite{Marti12}, and the stability of superfluid flow in a spinor condensate~\cite{Beattie13}.  
These experiments have given rise to theoretical studies discussing, e.g.,  the excitation spectrum and critical velocity of a superfluid BEC~\cite{Dubessy12}  and the simulation of the experiment~\cite{Ramanathan11} using the Gross-Pitaevskii equation~\cite{Mathey12,Piazza12} and the truncated Wigner approximation~\cite{Mathey12}. Most of the experimental and theoretical studies concentrate on the properties of persistent currents.   
The phase of a toroidal BEC changes by $2\pi k$ as the toroid is encircled, the integer $k$ being the winding number of the vortex. In a singly connected geometry a vortex with $|k|>1$ is typically unstable against splitting into
vortices with smaller $k$. In a multiply connected geometry 
 this process is suppressed for energetic reasons. 
In Ref.~\cite{Moulder12}  it was shown experimentally that a vortex with winding number three can persist in a toroidal single-component BEC for up to a minute. 
In other words, toroidal geometry makes it possible to avoid the fast vortex splitting taking place in a singly connected BEC and study the properties  of  vortices with large winding number. 
Instead of using a toroidal trap, a multiply connected geometry that stabilizes vortices can also be created by applying a Gaussian potential along the vortex core~\cite{Kuopanportti10}. 

In this paper, we calculate the Bogoliubov spectrum of a toroidal quasi-one-dimensional (1D) spin-1 BEC. Motivated by the experimental results of Refs.~\cite{Moulder12,Beattie13}, we assume that the splitting of vortices  occurs on a very long time scale in a spinor condensate where only one spin component is populated. The dominant instabilities can then be assumed to arise from the spin-spin interaction. For related theoretical studies on toroidal two-component condensates, see, for example, Refs.~\cite{Smyrnakis09,Anoshkin12}. 
In our analysis, the population of the $m_F=0$ spin component is taken to be zero initially, making it possible to calculate the excitation spectrum analytically. 
This type of a state can be prepared straightforwardly experimentally. The proliferation of instabilities can be observed by measuring the densities of the spin components.  

This paper is organized as follows. In Sec.~\ref{sec_Ham} we define the Hamiltonian, describe briefly the calculation of the excitation spectrum, and show that the spectrum can be divided into magnetization and spin modes. In Sec.~\ref{sec_Mag} we analyze the properties of the magnetization modes and 
illustrate how the presence of unstable modes can be seen experimentally. 
We also compare the analytical results with numerical calculations. 
In  Sec.~\ref{sec_Spin} we study the spin modes and their experimental observability analytically and numerically and show that a rotonlike spectrum can be realized both in rubidium and sodium condensates.  
In Sec.~\ref{sec_Exp} we discuss two recent experiments on toroidal BECs 
and show examples of the instabilities than can be realized in these systems.    
Finally, in Sec.~\ref{sec_Con} we summarize our results. 

\section{Energy and Hamiltonian}
\label{sec_Ham}
The order parameter of a spin-$1$ Bose-Einstein condensate reads $\psi=(\psi_{1},\psi_{0},\psi_{-1})^T$, 
where $T$ denotes the transpose. It fulfills the identity $\psi^\dag \psi =n_{3D}$, 
where $n_{3D}$ is the total particle density. 
We assume that the system is exposed to a homogeneous magnetic field oriented along the $z$ axis. 
The energy functional becomes, then, 
\begin{align}
\nonumber
&E[\psi]=\int d\mathbf{r} \left(\psi^\dag(\mathbf{r})\hat{H}_0(\mathbf{r})\psi(\mathbf{r})\right.\\
&\left. +\frac{1}{2}\left\{ g_0 n_{3D}^2(\mathbf{r}) + g_2 [\psi^\dag(\mathbf{r})\F\psi(\mathbf{r})]^2\right\}\right),
\label{eq_E}
\end{align}
where the single-particle Hamiltonian $\hat{H}_0$ is defined as
\begin{align}
\hat{H}_0(\mathbf{r})=-\frac{\hbar^2\nabla^2}{2m}+U(\mathbf{r})-\mu_{3D}-p\Fz+q\Fz^2, 
\end{align}
and $\F=(\Fx,\Fy,\Fz)$ is the (dimensionless) spin operator of a spin-1 particle, $U$ is the trapping potential, and $\mu_{3D}$ 
is the  chemical potential. The magnetic field introduces the linear and quadratic Zeeman terms, given by  $p$ and $q$, respectively. The sign of $q$ can be controlled experimentally   by using a linearly polarized microwave field \cite{Gerbier06}. 
The strength of the atom-atom interaction is characterized by $g_0=4\pi \hbar^2(a_0+2a_2)/3m$ and $g_2=4\pi \hbar^2(a_2-a_0)/3m$, where $a_F$ is the $s$-wave scattering length for two atoms colliding with total angular momentum $F$. 
The scattering lengths of ${}^{87}$Rb used here are $a_0=101.8a_B$ and $a_2=100.4 a_B$ \cite{vanKempen02}, measured in units of  the Bohr radius $a_B$.
For ${}^{23}$Na the corresponding values are $a_0=50.0a_B$ and $a_2=55.1a_B$ \cite{Crubellier99}.

The condensate is confined in a toroidal trap given in cylindrical coordinates as $U(r,z,\varphi)=m \left[\omega_r^2 (R-r)^2+\omega_z^2 z^2 \right]/2$,
where $R$ is the radius of the torus and $\omega_r,\omega_z$ are the trapping
frequencies in the radial and axial directions, respectively. We assume that the condensate is quasi-1D, so that the order parameter factors as 
$\psi(r,z,\varphi;t)=\psi_{r;z}(r,z)\psi_\varphi(\varphi;t)$, 
where $\psi_{r;z}$ is complex valued and time independent.
The normalization of $\psi_{r;z}$ is chosen such that $\int\int r dr dz |\psi_{r;z}(r,z)|^2=N/2\pi$, 
where $N$ is the total number of particles. This means that 
\begin{align}
\|\psi_\varphi(t)\|\equiv
\sqrt{\int_{0}^{2\pi}d\varphi\ \psi_\varphi^\dag(\varphi;t)\psi_\varphi(\varphi;t)}
\end{align}
has to be equal to $\sqrt{2\pi}$ for any $t$.
By integrating over $r$ and $z$ in Eq. \eqref{eq_E} we obtain
\begin{align}
\nonumber
& E_{1\textrm{D}}[\psi_\varphi] = \\
\nonumber
&\int_{0}^{2\pi}  d\varphi
\left( \psi_\varphi^\dag(\varphi)\left(-\epsilon \frac{\partial^2}{\partial \varphi^2} -\mueff-p\Fz+q\Fz^2\right)\psi_\varphi(\varphi)\right.\\
 &\left. +\frac{\neff}{2}\left\{ g_0 \left[\psi_\varphi^\dag(\varphi)\psi_\varphi(\varphi)\right]^2 + g_2 \left[\psi_\varphi^\dag(\varphi)\F\psi_\varphi(\varphi)\right]^2\right\}\right),
\label{eq_E1D}
\end{align}
where 
\begin{align}
\label{eq_epsilon}
\epsilon & =\frac{2\pi}{N}\frac{\hbar^2}{2m}\int_{0}^{\infty} rdr\int_{-\infty}^{\infty} dz\, \frac{1}{r^2} |\psi_{r;z}(r,z)|^2
\end{align} 
and 
\begin{align}
\label{eq_n}
n =\frac{2\pi}{N}\int_{0}^{\infty} rdr\int_{-\infty}^{\infty} dz\,|\psi_{r;z}(r,z)|^4.
\end{align}
In Eq. \eqref{eq_E1D} we have omitted an overall factor $N/2\pi$ multiplying the right-hand side of this equation.   
The chemical potential $\mu$ contains the original chemical potential $\mu_{3D}$ and terms coming from the integration of the kinetic and potential energies. 
 The magnetization in the $z$ direction, 
\begin{align}
f_z=\frac{1}{2\pi} \int_{0}^{2\pi} d\varphi\,\psi_\varphi^\dag (\varphi;t)\Fz\psi_\varphi (\varphi;t), 
\end{align} 
is a conserved quantity; the corresponding Lagrange multiplier can be included into $p$. 
In the following we drop the superscript $\varphi$ of $\psi_\varphi$.

We assume that in the initial state the spin is parallel to the magnetic field.
In \cite{Makela11} it was argued that in a homogeneous system the most unstable states are almost always of this form.
This state can be written as  
\begin{align}
\label{psipara}
\psi_\parallel(\varphi) = 
\frac{1}{\sqrt{2}}
\begin{pmatrix}
 e^{i k_1\varphi}\sqrt{1+f_z}\\
0 \\
e^{i\theta} e^{i k_{-1}\varphi}\sqrt{1-f_z}
\end{pmatrix},
\end{align}
where $\theta$ is the relative phase and 
the integer $k_{\pm 1}$ is the winding number of the $m_F=\pm 1$ component.  
The energy and stability of $\psi_\parallel$ are independent of $\theta$ and therefore we set $\theta=0$ in the rest of this article.    
If $k_1=1$ and $k_{-1}=0$, $\psi_\parallel$ describes a half-quantum vortex (Alice string), 
see, e.g., Refs. \cite{Leonhardt00,Isoshima01,Hoshi08}. 
The populations of $\psi_\parallel$ are time independent and the Hamiltonian giving the time evolution 
of $\psi_\parallel$ reads
\begin{align}
\label{Hparallel}
\HH_\parallel= \left(g_0 \neff - \mueff\right)\I
 +(g_2 \neff  f_z  -p_{\textrm{eff}})\Fz + q_{\textrm{eff}}\Fz^2,
\end{align} 
where 
\begin{align}
p_{\textrm{eff}}=& p-\frac{\epsilon}{2} (k_{1}^2-k_{-1}^2),\\
q_{\textrm{eff}}=& q+\frac{\epsilon}{2}(k_1^2+k_{-1}^2). 
\end{align}
The time evolution operator of $\psi_\parallel$ is $\hat{U}_\parallel(t)=e^{-it \HH_\parallel/\hbar}$.

We calculate the linear excitation spectrum in a basis where  $\psi_\parallel$ is stationary \cite{Makela11,Makela12} using the Bogoliubov approach, that is, we define 
 $\psi(\varphi;t)=\psi_\parallel(\varphi) +\delta\psi(\varphi;t)$ 
 and expand the time evolution equations to first order in $\delta\psi$. 
We write $\delta\psi=(\delta\psi_1,\delta\psi_0,\delta\psi_{-1})^T$ as  
\begin{align}
\label{eq_Bogoliubov}
\delta\psi_j(\varphi;t) \equiv e^{ik_j\varphi}\sum_{s=0}^{\infty} u_{j;s}(t)\,e^{i s \varphi}- v^*_{j;s}(t)\,e^{-i s\varphi},
\end{align}
where $j=0,\pm 1$ and $k_0\equiv 0$. Due to the toroidal geometry,  $\delta\psi_j(\varphi+2\pi;t)=\delta\psi_j(\varphi;t)$ 
has to hold. As a consequence, $s$ needs to be an integer. 
In the next two sections we analyze the excitation spectrum in detail; the actual calculation of the spectrum can be found in the appendix.   
The normalized wave function reads 
\begin{align}
\label{eq_psi}
\tilde{\psi}(\varphi;t) = c(t)[\psi_\parallel(\varphi)+\delta\psi(\varphi;t)],
\end{align}
where $c(t)$ is determined by the condition $\|\tilde{\psi}(t)\|=\sqrt{2\pi}$.
To characterize the eigenmodes we define 
\begin{align}
\label{eq_exp_Fz}
\langle\hat{F}_z\rangle (\varphi;t) \equiv \tilde{\psi}^\dag(\varphi;t)\hat{F}_z\tilde{\psi}(\varphi;t),
\end{align} 
so that $f_z=1/2\pi \int_0^{2\pi}d\varphi\ \langle\hat{F}_z\rangle(\varphi;t)$ for any $t$. 
Furthermore, we denote the population of the $m_F=0$ spin component by $\rho_0$, 
$\rho_0(\varphi;t)=|\tilde{\psi}_0(\varphi;t)|^2$. Note that here $\Fzexp$ and $\rho_0$ are calculated  in the basis where $\psi_\parallel$ is a stationary state. This basis and the original basis are related 
by a basis transformation that only affects the phases of the $m_F=\pm 1$ components. 
The densities of the spin components are thus identical in the original and new basis. The numerical calculations 
are done in the original basis.

The excitation spectrum can be divided into spin and magnetization modes.
The spin modes keep the value of $\langle\hat{F}_z\rangle$ unchanged in time, 
$\langle\hat{F}_z\rangle(\varphi;t)=\langle\hat{F}_z\rangle(\varphi;0)\approx f_z$, 
 but rotate the spin vector by making $\rho_0$ nonzero. The magnetization modes, on the other hand, lead to $\varphi$-dependent $\langle\hat{F}_z\rangle(\varphi;t)$,  
but leave $\rho_0$ unaffected. There are in total six eigenmodes. 
We denote them by $\hbar\omega_j$, where $j=1,2,3,4$ 
labels the magnetization modes and $j=5,6$ the spin modes. 
We denote the real and imaginary part of $\omega_{l}$ by $\omegar_{l}$ and 
$\omegai_{l}$, respectively. 
The mode labeled by $l$ is unstable if $\omegai_l$ is positive. 
We discuss first the magnetization modes.

\section{Magnetization modes}
\label{sec_Mag}
\subsection{Eigenmodes}
We characterize the eigenmodes by the quantities, 
\begin{align}
k_\pm =\frac{1}{2}\left(k_{1}\pm k_{-1}\right). 
\end{align}
Note that the value of $k_\pm$ can be a half-integer. 
The magnetization modes are independent of $q$ and can be written as
\begin{align}
\hbar\omega_l(s)=2\epsilon s k_{+}+\hbar\tilde{\omega}_l(s),  
\end{align}  
where $l=1,2,3,4$. 
The expression for $\tilde{\omega}_l$ is too long to be shown here. 
The value of $\tilde{\omega}_l$ depends on $k_{-}$ but is independent of $k_{+}$. 
Consequently, modes with differing $k_{+}$ but equal $k_{-}$ have identical stability. 

If $f_z=0$, the eigenvalues simplify and read 
\begin{align}
\nonumber
&\hbar\omega_{1,2,3,4}(s)\big|_{f_z=0} = 2\epsilon s k_{+}\\
&\pm \sqrt{\epsilon s^2\left[4\epsilon k_{-}^2+ w 
\pm \sqrt{16\epsilon k_{-}^2w  +(g_0 -g_2)^2 n^2}\right]},
\label{o1234fz0} 
\end{align}
where 
\begin{align}
w=\epsilon s^2+(g_0+ g_2)n. 
\end{align}
The signs are defined such that $++,-+,+-$, and $--$ correspond to $\omega_1,\omega_2,\omega_3$, and $\omega_4$, respectively.  
Unstable modes appear when the term inside the square brackets becomes negative. 
For rubidium and sodium $g_0+g_2 > 0$, which guarantees that $\omega_1$ and $\omega_2$ are real. Only $\omega_3$ can have a positive imaginary part. 
\begin{figure}[t]
\begin{center}
\includegraphics[scale=.92]{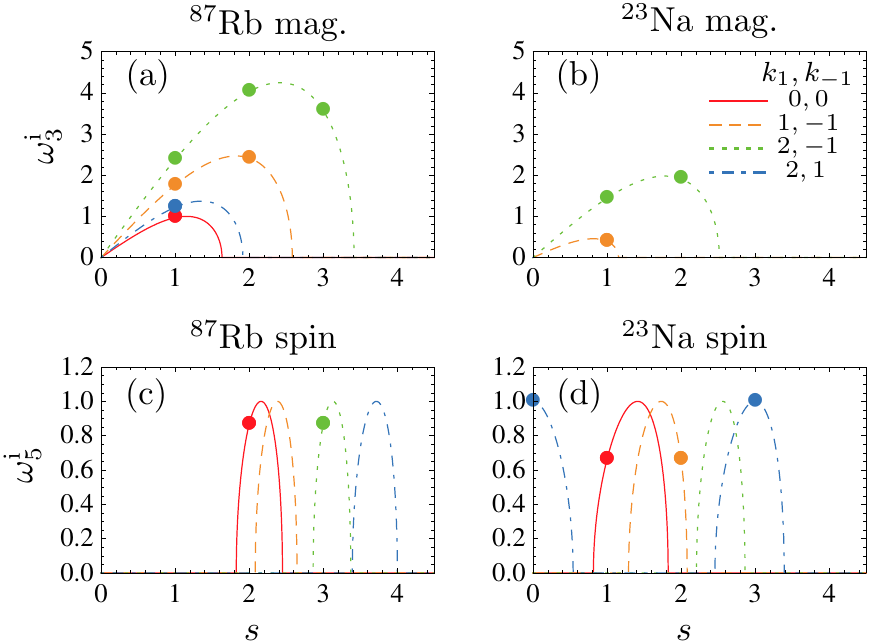}
\end{center}
\caption{(Color online) The amplitudes of the unstable spin and magnetization modes for rubidium and sodium. Here $\epsilon=0.75|g_2|n$, $q=2.5 |g_2|n$, $f_z=0$, and the unit of $\omegai_{3,5}$ is $|g_2|\neff/\hbar$. The lines have been drawn by treating $s$ as a continuous parameter; dots indicate the actual allowed nonvanishing values of $\omegai_{3,5}$. In (c) and (d) the curves are reflection symmetric with respect to 
$s=k_{+}=(k_{1}+k_{-1})/2$.      
\label{fig_wi}}
\end{figure}
As can be seen from Figs. \ref{fig_wi}(a) and 1(b), the value of $\omegai_3(s)$ grows as $|k_{-}|$ increases. The allowed values of $s$ are non-negative integers. The modes corresponding to $s=0$ are always stable, but unstable modes are present for $s= 1,2,\ldots, \lfloor\sqrt{4k_{-}^2-2 g_2 n/\epsilon}\rfloor$, where $\lfloor\cdots\rfloor$ is the floor function. 
Therefore, if there are $j$ unstable modes, they have to be the ones corresponding to $s=1,2,\ldots ,j$. 
 A lower bound for the value of $\epsilon$ yielding at least one  unstable mode is given by the equation $\epsilon (4k_{-}^2-1)\geq 2g_2 n$. In the case of a sodium BEC ($g_2 >0$) this means that the magnetization modes corresponding to  $k_{-}=0$ and $|k_{-}|=1/2$ are always stable. 
 This is visualized in Fig. \ref{fig_wi}(b), where $\omegai_3(s)$ corresponding to $(k_1,k_{-1})=(0,0)$ and  $(k_1,k_{-1})=(2,1)$ is seen to vanish for every $s$. 
In a rubidium condensate $(g_2<0)$ with $k_{-}=0$ unstable modes exist if $\epsilon\leq 2|g_2|n$; 
if $|k_{-}|>0$, instabilities are present regardless of the value of $\epsilon$.
For both rubidium and sodium the wave number $s$ of the fastest-growing instability is approximately given by the integer closest to $\sqrt{2/3}\sqrt{4k_{-}^2-2 g_2 n/\epsilon}$.  

\subsection{Experimental observability}
\begin{figure}[t]
\centering
\includegraphics[scale=0.85,clip]{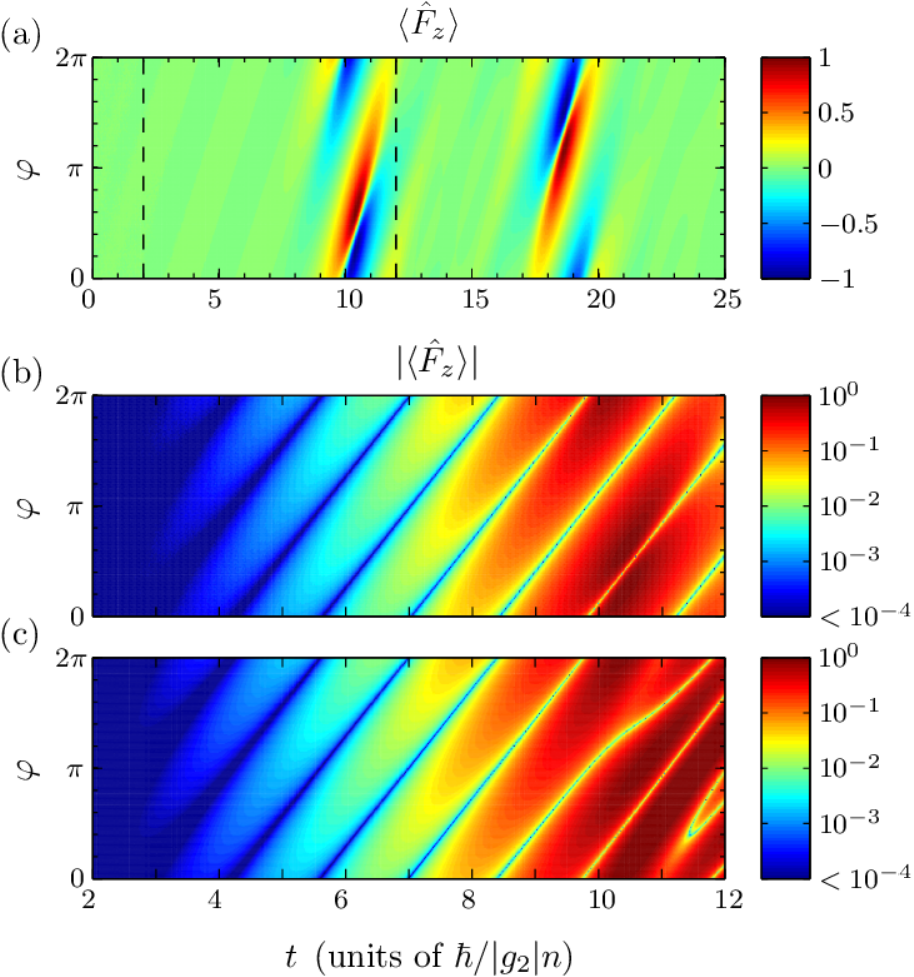}
\vspace{6mm}
\caption{(Color online) (a) Numerically calculated $\langle\hat{F}_z\rangle$ for the parameters corresponding to the blue dash-dotted line in Fig.~\ref{fig_wi}(a), that is, a ${}^{87}$Rb condensate 
with $\epsilon =0.75 |g_2|n, q=2.5|g_2|n, f_z=0, k_{1}=2,$ and $k_{-1}=1$. (b) Magnification of the region bounded by the dashed vertical lines in (a). Here we plot $|\Fzexp|$ instead of $\Fzexp$ and use a logarithmic scale to make the initial growth of $|\langle\hat{F}_z\rangle|$ visible. (c) Analytically calculated $\langle\hat{F}_z\rangle$, see Eq.~\eqref{eq_FzexpApprox}. 
\label{fig_num_mag}
}
\end{figure}

The properties of unstable magnetization modes can be studied experimentally 
by measuring $\Fzexp$. We assume that there is one dominant unstable mode and that $f_z=0$. 
The initial time evolution of $\langle\hat{F}_z\rangle$ reads, then (see the appendix), 
\begin{align}
\label{eq_FzexpApprox}
\nonumber
&\langle \hat{F}_z\rangle (\varphi;t) \approx c^2(t)\left\{ A  e^{\omegai_3 t} 
\cos\left[\theta+s\left(\varphi-\frac{2\epsilon k_{+} t}{\hbar}\right)\right]\right.\\
&\left.+ B e^{2\omegai_3 t} \cos\left[2\theta+2s\left(\varphi-\frac{2\epsilon k_{+} t}{\hbar}\right)\right]\right\},
\end{align}
where $c$ is the normalization factor appearing in Eq.~\eqref{eq_psi} and $A,B,$  and $\theta$ are defined in Eqs.~\eqref{eq_A}, \eqref{eq_B},  and \eqref{eq_theta}, respectively. Because typically  
$B\ll A$, the first term on the right-hand side  
of Eq.~\eqref{eq_FzexpApprox} dominates over the second term during the initial time evolution. 
This leads to $\langle\hat{F}_z\rangle$ having $s$ maxima and minima.  
If $\kp\not =0$, these maximum and minimum regions rotate around the torus as time evolves, indicating that the behavior of  $\Fzexp$ depends on $\kp$, even though the growth rate of the instabilities $\omegai_3$ is independent of $\kp$. We study the validity of Eq.~\eqref{eq_FzexpApprox} by considering a rubidium condensate with $\epsilon=0.75|g_2|n$, 
$q=2.5|g_2|n$, $k_{1}=2$, and $k_{-1}=1$, corresponding to the blue dash-dotted line in Figs.~\ref{fig_wi}(a) and~\ref{fig_wi}(c). 
Analytical results predict that the only unstable mode of this system is a magnetization mode corresponding to $s=1$. The numerically calculated time evolution of $\langle\hat{F}_z\rangle$ is shown  
in Figs.~\ref{fig_num_mag}(a) and \ref{fig_num_mag}(b).  
The $s=1$ magnetization mode can be seen to be unstable. 
The rotation of the minimum and maximum of $\Fzexp$ around the torus is clearly visible in Fig.~\ref{fig_num_mag}. The analytically obtained behavior of $\langle\hat{F}_z\rangle$ is shown in Fig. \ref{fig_num_mag}(c). 
By comparing  Figs.~\ref{fig_num_mag}(b) and  \ref{fig_num_mag}(c), we see that Eq.~\eqref{eq_FzexpApprox} describes the time evolution of $\langle\hat{F}_z\rangle$ very precisely 
up to $t\approx 10\hbar/|g_2|n$.  
The only parameters in Eq.~\eqref{eq_FzexpApprox} that are not fixed by the parameters 
used in the numerical calculation are the initial global phase and length 
$\|\delta\psi(t=0)\|$ of $\delta\psi(t=0)$.   
In Fig.~\ref{fig_num_mag}(c) we have chosen the values of these variables in such a way that the match between the numerical and analytical results is the best possible.

\section{Spin modes}
\label{sec_Spin}
\subsection{Eigenmodes} 
We now turn to the spin modes. As shown in the appendix, the spin modes read
\begin{align}
\label{o56}
&\hbar\omega_{5,6}(s) =2\epsilon k_{+}(s-k_{+})\\
\nonumber 
&\pm \sqrt{\left\{\epsilon [(s-k_+)^2-k_{-}^2]+g_2 \neff-q\right\}^2-(1-f_z^2)(g_2 \neff)^2},
\end{align}
where $+$ ($-$) corresponds to $\omega_5$ ($\omega_6$). 
If $k_{+}=0$, the effect of vortices  can be taken into account by scaling $q\rightarrow q+\epsilon k_{-}^2$, {\it i.e.}, the spin modes of a system with $(k_1,k_{-1})=(k,-k)$ and $q=\tilde{q}$  are equal to the spin modes of a vortex-free condensate with $q=\tilde{q}+\epsilon k^2$. Spin modes are unstable if and only if the term inside the square root is negative. Now only $\omega_5$ can have a positive imaginary part.   
The fastest-growing unstable mode is obtained at $\epsilon [(s-k_+)^2-k_{-}^2]+g_2 \neff-q=0$ and has the amplitude $\hbar\omegai_5(s)=|g_2|n\sqrt{1-f_z^2}$. 
Unlike in the case of the magnetization modes, the maximal amplitude is bounded from above and is independent of the winding numbers [see Figs.~\ref{fig_wi}(c) and \ref{fig_wi}(d)]. By adjusting  the strength of the magnetic field, the fastest-growing unstable mode can be chosen to be located at a specific value of $s$, showing that it is 
easy to adjust the stability properties experimentally. 
At $f_z=0$ the width of the region on the $s$-axis giving positive $\omegai_5$ is 
$|\sqrt{k_{-}^2+q/\epsilon}-\sqrt{k_{-}^2+q/\epsilon- 2g_2 n/\epsilon}|$. 
This region can thus be made narrower by increasing $\epsilon,k_{-}$, or $q$. 
Since the magnetization modes are insensitive to the magnetic field, the properties of the spin and magnetization modes can be tuned independently.  
The winding number dependence of unstable spin modes is illustrated 
in Figs. \ref{fig_wi}(c) and \ref{fig_wi}(d). 

\subsection{Rotonlike spectrum}
\begin{figure}[t]
\begin{center}
\includegraphics[scale=.90]{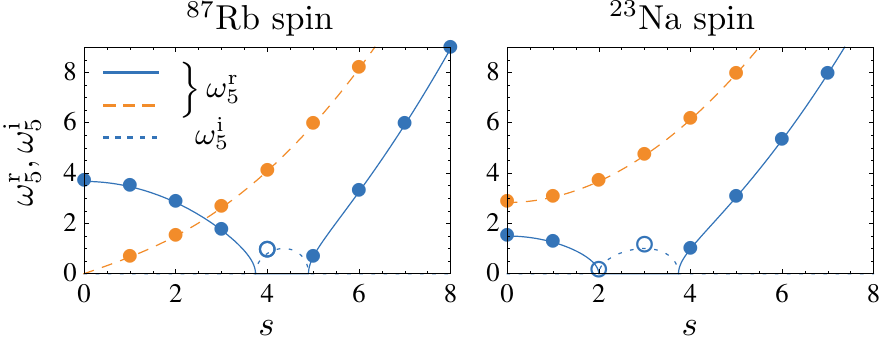}
\end{center}
\caption{(Color online) The real ($\omegar_5$) and imaginary ($\omegai_5$) component of the spin mode $\omega_5$ for rubidium and sodium. Here $\epsilon=0.2|g_2|n, f_z=0, k_{1}=-k_{-1}$, and $k_1$ is an arbitrary integer. For the blue solid and blue dotted lines $q+\epsilon k_{-}^2 =2.8|g_2|n$ and for the orange dashed line 
$q+\epsilon\km^2 =-2|g_2|n$. The unit of $\omega_5^{\textrm{r},\textrm{i}}$ is $|g_2|\neff/\hbar$.  
The lines have been drawn by treating $s$ as a continuous parameter; dots  (open circles) indicate the actual allowed nonvanishing values of $\omegar_5$ ($\omegai_5$). 
\label{fig_roton}}
\end{figure}

Interestingly, by tuning $\epsilon$ and $q$,  a rotonlike spectrum can be realized (see the solid and dotted blue lines in Fig. \ref{fig_roton}). 
Now the phonon part of the spectrum is missing, but the roton-maxon feature is present. For $f_z=k_{+}=0$, the roton spectrum exists if  
$q\geq \max\{0,2g_2 n\}$. Because only integer values of $s$ are allowed, 
it may happen that $\omegai_{5}$ is nonzero only in some interval of the $s$ axis that does not contain integers [see Figs. \ref{fig_wi}(c) and \ref{fig_wi}(d) for examples of this in the context of magnetization modes]. 
In this case the rotonic excitations are stable. Alternatively, there can be unstable modes close to the roton minimum  (see Fig.~\ref{fig_roton} and Ref.~\cite{Matuszewski10}).   
As evidenced by the orange dashed lines in Fig. \ref{fig_roton}, the roton spectrum can be made to vanish simply by decreasing $q$. Also the values of $s$ leading to unstable modes can be controlled by varying $q$. For example, using the parameter values corresponding to the blue solid line in Fig. \ref{fig_roton}, we find that by decreasing (increasing) the value of $q+\epsilon\km^2$ from $2.8|g_2| n$ to $|g_2| n$ ($4|g_2| n$), the $s=3$ ($s=5$) mode can be made unstable in a rubidium condensate. This opens the way for quench experiments of the type described in Refs.~\cite{Sadler06,Bookjans11}.  
Instead of altering $q$, instabilities can also be induced by making $\epsilon$ smaller by changing the trapping frequencies. It is known that a rotonlike spectrum can exist in various types of BECs, such as  in a dipolar condensate  (see, e.g., Refs.~\cite{Odell03,Santos03,Cherng09}), in a Rydberg-excited condensate~\cite{Henkel10}, or in a spin-1 sodium condensate prepared in a specific state~\cite{Matuszewski10}. 
In the present case the rotonlike spectrum exists both in a sodium and rubidium BEC and the state 
[Eq.~\eqref{psipara}] giving rise to it is easy to prepare experimentally. 
Note that the roton-maxon feature exists also in a vortex-free condensate and for 
any $|f_z|<1$. These results suggest that the roton-maxon character of the spectrum is rather a rule than an exception in spinor BECs.

\subsection{Experimental observability}
\begin{figure}[ht]
\centering
\includegraphics[scale=0.85,clip]{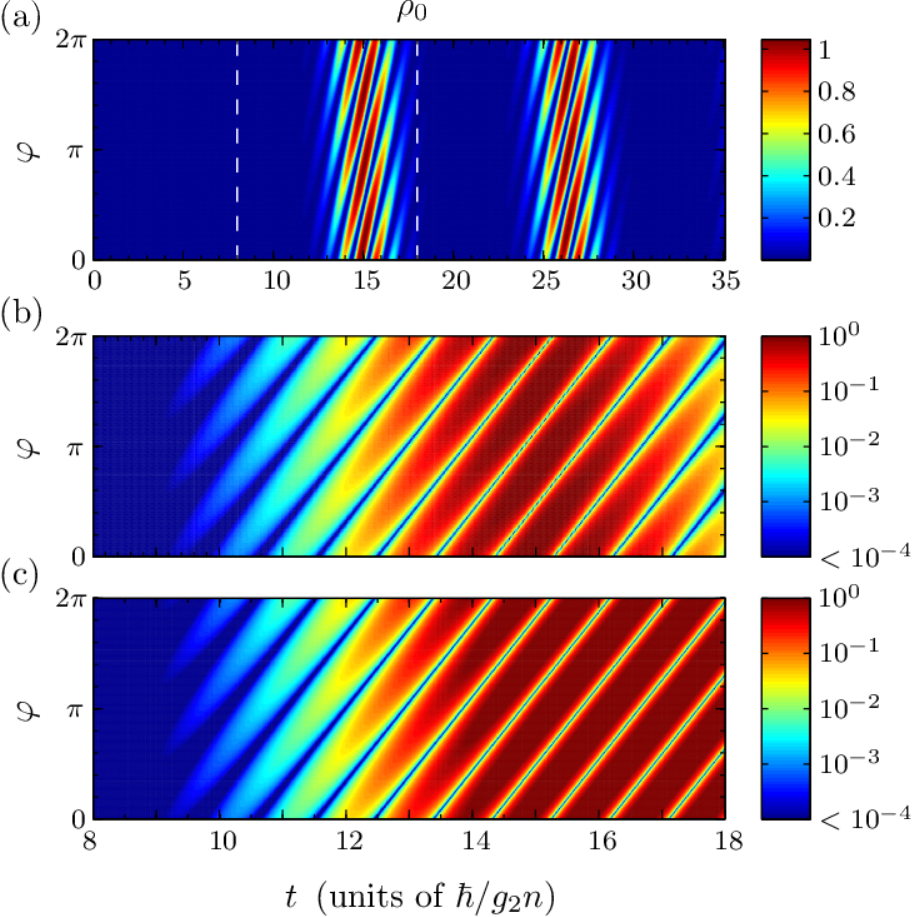}
\vspace{5mm}
\caption{(Color online) (a) Numerically calculated $\rho_0$ for a ${}^{23}$Na condensate 
with $\epsilon =0.75 g_2n, q=2.5g_2n, f_z=0, k_{1}=2,$ and $k_{-1}=1$, corresponding to the 
blue dash-dotted line in Fig.~\ref{fig_wi}(d). (b) A magnification of the region bounded by the dashed vertical lines in (a). (c) Analytically calculated $\rho_0$.  
In (b) and (c) a logarithmic scale has been used. 
\label{fig_num_spin} 
}
\end{figure}

The properties of unstable spin modes can be studied experimentally by measuring $\rho_0$. 
Assuming that there is one dominant unstable spin mode located at wave number $s$, we find that (see the Appendix)
\begin{align}
&\delta\psi_0(\varphi;t)\propto e^{i k_+\varphi + \omegai_5 t}
\sin\left[\left(s-k_{+}\right)\left(\varphi-\frac{2\epsilon\kp t}{\hbar}\right)+\frac{\thetaSpin}{2}\right].
\label{eq_deltapsi0}
\end{align}
The phase $\thetaSpin$ is defined in Eq.~\eqref{eq_thetaSpin}. 
The sign of $\delta\psi_0$ changes at every point where the density $\rho_0\propto |\delta\psi_0|^2$ vanishes. This is similar to the behavior of the phase of a dark soliton \cite{Frantzeskakis10}. 
The number of nodes in $\rho_0$ is  $2|s-k_{+}|$, that is,  
if $2k_{+}$ is even (odd), $\rho_0$ has an even (odd) number of nodes.
The density peaks resulting from the instability rotate around the torus if $k_{+}(s-k_{+})$ is  nonzero. In the special case $s=\kp$ the density $\rho_0(\varphi;t)$ is independent of 
$\varphi$. A numerically obtained example of this is shown in Fig.~\ref{fig_Ramanathan}(a). 
In Fig.~\ref{fig_num_spin} we compare numerical calculations to analytical results. 
We consider a sodium condensate with $\epsilon=0.75 g_2 n,q=2.5g_2n,k_{1}=2$, and $k_{-1}=1$.  
For these values the $s=3$ spin mode is the only unstable mode [see the blue dash-dotted line in Figs.~\ref{fig_wi}(b) and \ref{fig_wi}(d)]. Numerical calculations give the same result. 
By comparing Figs.~\ref{fig_num_spin}(b) and \ref{fig_num_spin}(c) we see that the analytical 
expression for $\rho_0$ approximates the actual dynamics very precisely up to $t\approx 15\hbar/g_2n$. 
As in the case of the magnetization modes, we choose the initial length and overall phase of $\delta\psi(t=0)$ in such a way that the agreement between the numerical and analytical results is the best possible.

\section{Experiments}
\label{sec_Exp}
In this section we calculate the ratio $\epsilon/|g_2|n$ corresponding to two recent experiments. 
To obtain an analytical estimate for $\epsilon$, we assume that the particle density $|\psi_{r;z}(r,z)|^2$ is peaked around $R$ and approximate $1/r^2\approx 1/R^2$ in Eq.~\eqref{eq_epsilon}. This gives $\epsilon\approx \hbar^2/2mR^2$.  
Approximating $\psi_{r;z}$ by the Thomas-Fermi (TF) wavefunction yields 
\begin{align}
n &\approx\sqrt{\frac{2 m N\omega_r\omega_z}{9\pi^2g_0 R}}.
\end{align} 
We see that $\epsilon/|g_2|n \propto (\omega_r\omega_z N R^3)^{-1/2}$, so that the properties of the excitation spectrum can be controlled by adjusting the trapping frequencies, number of particles, and the radius of the toroid.

\begin{figure}
\center
\includegraphics[scale=0.85,clip]{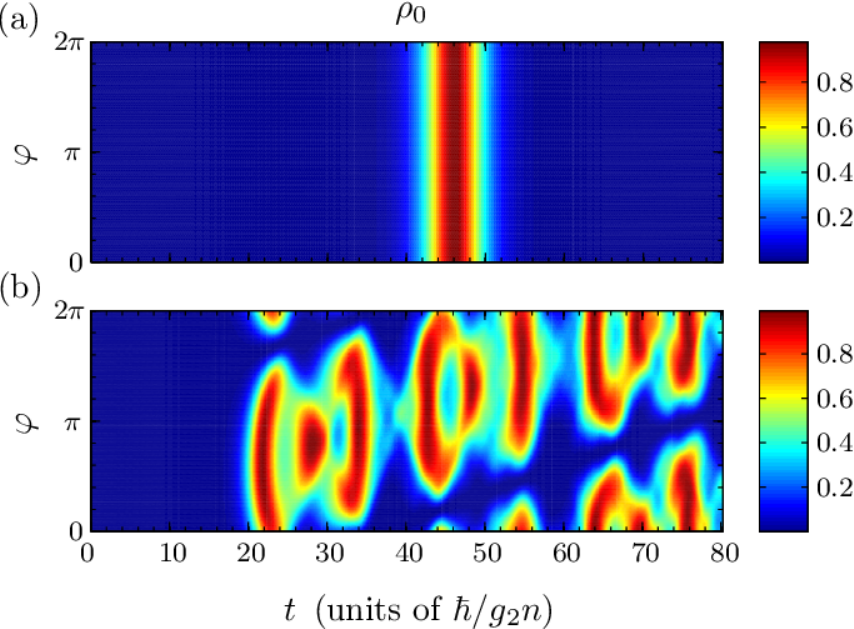}
\vspace{5mm}
\caption{(Color online) Numerically calculated $\rho_0$ for a ${}^{23}$Na condensate with $\epsilon=q=0.04|g_2|n$ and $f_z=0$. In (a) $k_{1}=k_{-1}=0$ and in 
(b) $k_{1}=1,k_{-1}=0$. The value of $\epsilon$ corresponds to that of \cite{Ramanathan11}.  
\label{fig_Ramanathan}}
\end{figure}

Using the parameter values of the sodium experiment \cite{Ramanathan11} we get $\epsilon\approx 0.04 g_2 n$. We study numerically the cases  $(k_{1},k_{-1})=(0,0)$ and  $(k_{1},k_{-1})=(1,0)$.
With the help of Eqs.~\eqref{o1234fz0} and \eqref{o56} we find that magnetization modes are stable, 
but spin modes are unstable in both cases. If $0< q \leq 0.04 g_2 n$, $f_z=0$, and  $(k_{1},k_{-1})=(0,0)$, the unstable spin mode leads to a position-independent, homogeneous, increase in  $\rho_0$.  If  $(k_{1},k_{-1})=(1,0)$, we get $\rho_0(\varphi;t)\sim e^{2\omegai_5 t}\sin^2[(\epsilon t+\varphi)/2]$. The $1$D numerical calculations 
shown in Fig. \ref{fig_Ramanathan} confirm the validity of these analytical predictions. 
This example illustrates that even a small $\epsilon$ can lead to a strongly winding number-dependent 
behavior of $\rho_0$. 

\begin{figure}
\center
\includegraphics[scale=0.85,clip]{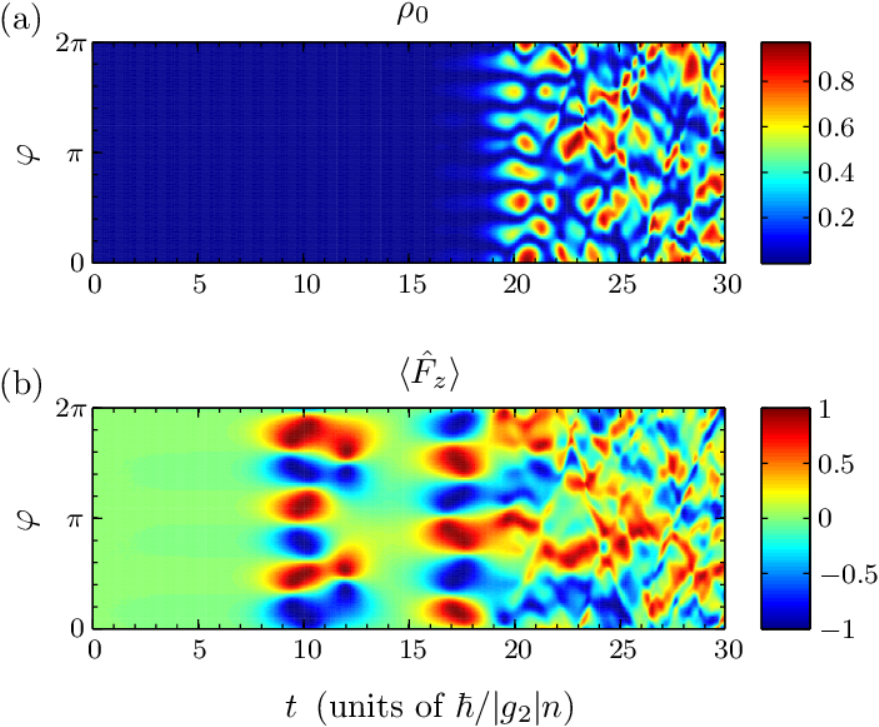}
\vspace{5mm}
\caption{(Color online) Numerically calculated (a) $\rho_0$ and (b) $\Fzexp$ for
 a ${}^{87}$Rb condensate with $\epsilon=0.2|g_2|n,q=2.6|g_2|n,f_z=0$, and $k_{1}=-k_{-1}=1$. 
 The value of $\epsilon$ corresponds to that of Ref.~\cite{Beattie13}.
\label{fig_Beattie}}
\end{figure}

The first experimental realization of a toroidal spin-1  
BEC was reported recently~\cite{Beattie13}. The stability of a rubidium BEC with a winding number three vortex  in the $m_F=1$ and $m_F=0$ components was found to depend strongly on the population difference of the two components, the most unstable situation corresponding to equal population. Although not directly comparable, our analysis agrees qualitatively with this result: The growth rate of unstable spin and magnetization  modes increases as the population difference of the $m_F=1$ and $m_F=-1$ components goes to zero. 
The parameter values of this experiment  yield $\epsilon\approx 0.20 |g_2|n$. 
The $s=1,2,$ and $s=3$ magnetization modes are unstable regardless of the values of winding numbers. 
 If $k_{+}=0$ and $q+\epsilon\km^2 = 2.8|g_2|n$, the spin modes have a rotonlike spectrum (see the left panel of Fig.~\ref{fig_roton}). The $s=4$  mode can be seen to be the only unstable spin mode. 
This is confirmed by the numerical results shown in Fig.~\ref{fig_Beattie}(a). In this figure we have chosen 
$k_{1}=-k_{-1}=1$ and $q=2.6|g_2|n$, so  that $q+\epsilon\km^2 =2.8|g_2|n$. 
Because $\kp=0$, Eqs.~\eqref{eq_FzexpApprox} and \eqref{eq_deltapsi0} predict that the nodes of $\rho_0$ and $\Fzexp$ do not rotate around the torus as time evolves. This is clearly the case in Fig.~\ref{fig_Beattie}.
The $s=3$ magnetization mode can be seen to be the fastest growing unstable mode. However,  around $t\approx 12\hbar/g_2 n$, the $s=2$ mode becomes the dominant unstable mode. These observations agree with analytical predictions: Using Eq.~\eqref{o1234fz0} we find that $\hbar\omegai_3(s)/|g_2|n=0.72, 1.26$, and $1.34$ for $s=1,2$, and $s=3$, respectively. For other values of $s$ we get $\omegai_3(s)=0$.    

\section{Conclusions}
\label{sec_Con}
We have calculated analytically the Bogoliubov spectrum of a toroidal spin-1 BEC that has vortices in the $m_F=\pm 1$ spin components and is subjected to a homogeneous magnetic field. 
We treated the strength of the magnetic field and the winding numbers of the vortices as free parameters and assumed that the population of the $m_F=0$ component vanishes. We assumed also that the system is  quasi-one-dimensional. We found that the spectrum can be divided into spin and magnetization modes. Spin modes 
change the particle density of the $m_F=0$ component but leave the particle density difference of the $m_F=1$ and $m_F=-1$ components unchanged. The magnetization modes do the opposite. 
An important parameter characterizing the spectrum is the ratio of the kinetic to interaction energy, $\epsilon/|g_2|n$. 
The properties of magnetization modes can be tuned by adjusting this ratio, whereas in the case of spin modes also the strength of the magnetic field can be used to control the spectrum. For example, a spin mode spectrum with a roton-maxon structure can be realized both in rubidium and sodium condensates by making the magnetic field strong enough. Furthermore, by changing the strength of the magnetic field or the ratio $\epsilon/|g_2|n$, an initially stable condensate can be made unstable. We also showed that some unstable spin modes lead to a transient dark solitonlike wave function of the $m_F=0$ spin component. 
Finally, we discussed briefly two recent experiments on toroidal BECs and 
 showed examples of the instabilities that can be realized in these systems.  

We studied the validity of the analytical results by numerical one-dimensional simulations, finding that the former give a very good description of the stability of the condensate and the initial time evolution of the instabilities. 

\begin{acknowledgments}
This research has been supported by the Alfred \mbox{Kordelin} Foundation and the Academy of Finland through
its Centres of Excellence Program (Project No. 251748).
\end{acknowledgments}

\appendix

\section{Calculation of the excitation spectrum}
\label{appendix_calculation}
Following Refs. \cite{Makela11,Makela12},  we calculate the excitation spectrum in a basis where  $\psi_\parallel$ is stationary. This basis can be defined easily because the time evolution operator 
$\hat{U}_\parallel(t)=e^{-it \HH_\parallel/\hbar}$ is known.  
In this basis, the energy of an arbitrary state $\psi$ is given by
\begin{align}
E_{1\textrm{D}}^{\textrm{new}}[\psi]\equiv E_{1\textrm{D}}[\U_\parallel\psi]
+i\hbar\langle\psi |\left(\frac{\partial}{\partial t}\U^{-1}_\parallel\right)\U_\parallel\psi\rangle,
\end{align}
and the time evolution of $\psi=(\psi_1,\psi_0,\psi_{-1})^T$ can be obtained from  
\begin{align}
\label{eq_DE}
i\hbar\frac{\partial}{\partial t} \psi_m =\frac{\delta E_{1\textrm{D}}^{\textrm{new}}[\psi]}{\delta\psi_m^*},\quad m=-1,0,1.
\end{align}
Here $T$ denotes the transpose and $*$ the complex conjugate. 
We write the (unnormalized) wavefunction in the new basis as $\psi(\varphi;t)=\psi_\parallel(\varphi) +\delta\psi(\varphi;t)$, where  the components of $\delta\psi$ read 
\begin{align}
\label{eq_deltapsi}
\delta\psi_m(\varphi;t) \equiv e^{ik_m\varphi}\sum_{s=0}^{\infty} u_{m;s}(t)\,e^{i s \varphi}- v^*_{m;s}(t)\,e^{-i s\varphi}.
\end{align}
Here $m=0,\pm 1$ and $k_0\equiv 0$. By expanding Eq.~\eqref{eq_DE} to first order in $\delta\psi$ and using Eq.~ \eqref{eq_deltapsi} we get 
the equations,
\begin{align}
i\hbar\frac{\partial}{\partial t} 
\begin{pmatrix} 
u_{0;s}(t) \\
 v_{0;s-2k_{+}}(t)
\end{pmatrix}
=\Btwo(t)
\begin{pmatrix} 
u_{0;s}(t) \\
 v_{0;s-2k_{+}}(t)
\end{pmatrix},
\end{align}
and 
\begin{align}
i\hbar\frac{\partial}{\partial t} 
\begin{pmatrix}
u_{1;s}(t)\\
u_{-1;s}(t)\\ 
v_{1;s}(t)\\
v_{-1;s}(t). 
\end{pmatrix}
=
\Bfour
\begin{pmatrix}
u_{1;s}(t)\\
u_{-1;s}(t)\\ 
v_{1;s}(t)\\
v_{-1;s}(t)
\end{pmatrix}.
\end{align}
The $\Btwo$ matrix reads
\begin{align}
\Btwo(t) =  
\begin{pmatrix}
\epsilon s^2+ g_2 n & -e^{-i\alpha t} g_2 n \sqrt{1-f_z^2}\\
-e^{i\alpha t} g_2 n \sqrt{1-f_z^2} & 
\epsilon (s-2k_+)^2+ g_2 n
\end{pmatrix},
\end{align}
where 
\begin{align}
\alpha=\frac{\epsilon(k_1^2+k_{-1}^2)+2 q}{\hbar}.
\label{eq_alpha}
\end{align}
$\Bfour$ can be written as
\begin{align}
\Bfour =
\begin{pmatrix}
\epsilon s^2 \hat{\mathbb{I}}_2+\hat{X}+\hat{D} & -\hat{X} \\
\hat{X} & -\epsilon s^2 \hat{\mathbb{I}}_2-\hat{X}+\hat{D}
\end{pmatrix},
\end{align}
where $\hat{\mathbb{I}}_2$ is the $2\times 2$ identity matrix,  $\hat{D}= 2 \epsilon s\, \textrm{diag}(k_{1},k_{-1})$, and $\hat{X}$ is defined as
\begin{align}
\hat{X}=\frac{n}{2}
\begin{pmatrix}
(g_0+g_2)(1+f_z) & (g_0-g_2)\sqrt{1-f_z^2}\\
(g_0-g_2)\sqrt{1-f_z^2} & (g_0+g_2)(1-f_z)
\end{pmatrix}.
\end{align}
The eigenvalues of $\Btwo$ and $\Bfour$ give the spin and magnetization modes, respectively. 
We write the eigenvalues of these matrices as $\hbar\omega_j$, where $j=1,2,3,4$ 
labels the magnetization modes and $j=5,6$ labels the spin modes. 
We write the wave function $\delta\psi$ as 
\begin{align}
\delta\psi(\varphi;t)=\sum_{j=1}^{6}\delta\psi^j(\varphi;t),
\end{align}
where $\delta\psi^j=(\delta\psi_1^j,\delta\psi_0^j,\delta\psi_{-1}^j)^T$ and 
\begin{align}
\label{eq_deltapsij}
\delta\psi_m^j(\varphi;t) = e^{ik_m\varphi}\sum_{s=0}^{\infty} u_{m;s}^j(t)\,e^{i s \varphi}
-v^{j*}_{m;s} (t)\,e^{-i s\varphi},
\end{align}
$m=-1,0,1,j=1,2,3,4,5,6$. Here $u^j$ and $v^j$ are written in terms of the 
eigenvector of $\Bfour$ or $\Btwo$ corresponding to the eigenvalue $\hbar\omega_j$.

\subsection{Eigenvalues and eigenvectors of $\Bfour$}
The eigenvalues of $\Bfour$ for a general value of $f_z$ can be calculated straightforwardly but they are too long to be shown here. The eigenvalues at $f_z=0$  are given in Eq. \eqref{o1234fz0}. 
The wave function $\delta\psi^j$  is of the form $\delta\psi^j =(\delta\psi_{1}^j,0,\delta\psi_{-1}^j)^T$, 
where 
\begin{align}
\label{eq_deltapsi_mag}
&\delta\psi_{\pm 1}^j (\varphi;t) = \sum_{s=0}^{\infty} \left[u_{\pm 1;s}^j(t)
\,e^{is\varphi}
- v^{j*}_{\pm 1;s}(t)\,e^{-is\varphi}\right], 
\end{align}
$j=1,2,3,4$. As is the case with the eigenvalues of $\Bfour$, 
for a general value of $f_z$ the eigenvectors are very complex. 
We therefore set $f_z=0$ in the following. Furthermore, we only calculate the eigenvector 
corresponding to the eigenvalue $\hbar\omega_3$, which is the only eigenvalue 
that can have a positive imaginary part.  We assume that there is a dominant instability at wavenumber $s$, so that $\omegai_3(s)>0$.  
The corresponding eigenvector reads 
\begin{align}
\label{eq_w1m1}
\begin{pmatrix}
u_{1;s}^3(0)\\
u_{-1;s}^3(0)\\
v_{1;s}^3(0)\\
v_{-1;s}^3(0)
\end{pmatrix}
=
\absh e^{i\deltah}
\begin{pmatrix}
e^{i(\delta+\deltac)}\\
e^{i\deltac}\absc\\
e^{i\delta} \absc\\
1 
\end{pmatrix},
\end{align} 
where $h=\absh e^{i\deltah}$ determines the length and overall phase of the eigenvector and 
\begin{align}
&\cc=\absc e^{i\deltac} = \frac{i \hbar\omega_3^\textrm{i}+\epsilon s(2k_{-}+s)}{i \hbar\omega_3^\textrm{i}+\epsilon s(2k_{-}-s)},\\ 
\nonumber
\label{eq_delta}
&\delta =\arg\big\{\epsilon s^2[4 g_0 g_2 n^2+\epsilon s^2 (g_0+g_2)n]\\
& -(g_0+g_2)n(2\epsilon s k_{-} +i\hbar \omegai_3)^2\big\}.
\end{align}
If $k_{1}=k_{-1}$, Eq.~\eqref{eq_delta} becomes $\delta=\arg((g_0-g_2)g_2)$.
With the help of Eqs.~\eqref{eq_deltapsi_mag}--\eqref{eq_delta} we obtain 
\begin{align}
\label{eq_Fzexp}
\nonumber
&\psi^\dag(\varphi;t)\hat{F}_z\psi(\varphi;t)=A  e^{\omegai_3 t} \cos\left(\theta+s\varphi-\omegar_3 t\right)\\ 
& +B e^{2\omegai_3 t} \sin\left[2\left(\theta+s\varphi-\omegar_3 t\right)\right],
\end{align}
where
\begin{align}
\label{eq_A}
A &=-2\sqrt{2}\absh \left[\absc\cos\left(\frac{\delta-\deltac}{2}\right)-\cos\left(\frac{\delta+\deltac}{2}\right)\right],\\
\label{eq_B}
B &= 2\absh^2 \absc \sin(\delta),
\end{align}
and
\begin{align}
\label{eq_theta}
\thetaMag &= \frac{\delta+\deltac+2\deltah}{2}.
\end{align}

\subsection{Eigenvalues and eigenvectors of $\Btwo$}
In the case of the spin modes $\delta\psi^j=(0,\delta\psi^j_0,0)^T$, where  
\begin{align}
\label{eq_deltapsi0Appendix}
\delta\psi_0^j (\varphi;t) =  \sum_{s=0}^{\infty} \left[ u_{0;s}(t)\,e^{is\varphi}
-v^*_{0;s-2 k_+}(t)\,e^{-i(s-2k_+)\varphi}\right],
\end{align}
$j=5,6$. The time dependence of $\Btwo$ can be eliminated by defining 
a new basis as 
\begin{align} 
\begin{pmatrix}
\tilde{u}_{0;s}(t)\\
\tilde{v}_{0;s-2\kp}(t)
\end{pmatrix}
=U(t)
\begin{pmatrix}
u_{0;s}(t)\\
v_{0;s-2\kp}(t)
\end{pmatrix},
\end{align}
where 
\begin{align}
U(t)= 
\begin{pmatrix}
0 & e^{-\frac{i \alpha t}{2}}  \\
e^{\frac{i \alpha t}{2}}  & 0
\end{pmatrix},
\end{align}
and $\alpha$ is defined in Eq.~\eqref{eq_alpha}. In the new basis the time evolution is determined by 
the operator
\begin{align}
\hat{\tilde{B}}_2 = U(t)\Btwo(t)U^\dag (t)+i\hbar \left[\frac{d}{dt}U(t)\right]U^\dag (t),
\end{align}
which is time independent. The eigenvalues of $\hat{\tilde{B}}_2$ are 
\begin{align}
&\hbar\omega_{j}(s)=2\epsilon k_{+}(s-k_{+})- (-1)^j\sqrt{a^2-b^2},
 \end{align}
where $j=5,6$, and we have defined 
\begin{align}
a &= \epsilon[(s-k_+)^2-k_-^2]+g_2 n-q,\\
b & = \sqrt{1-f_z^2} g_2 n. 
\end{align} 
The eigenvector corresponding to $\hbar\omega_j(s)$ reads 
\begin{align} 
\begin{pmatrix}
\tilde{u}_{0;s}^j(0)\\
\tilde{v}_{0;s-2\kp}^j(0)
\end{pmatrix}
=\h 
\begin{pmatrix}
\frac{a+(-1)^j \sqrt{a^2-b^2}}{b}\\
1
\end{pmatrix},  
\end{align}
where $\h$ is an arbitrary nonzero complex number. This gives 
\begin{align}
\begin{pmatrix}
u_{0;s}^j(t)\\
v_{0;s-2\kp}^j(t)
\end{pmatrix}
=
he^{-i\omega_j t}\begin{pmatrix}
e^{-\frac{i\alpha t}{2}}\\
e^{\frac{i\alpha t}{2}}\frac{a+(-1)^j \sqrt{a^2-b^2}}{b}
\end{pmatrix}.
\end{align} 
Using Eq.~\eqref{eq_deltapsi0Appendix} we get
\begin{align}
\label{eq_deltapsi0Appendix2}
\nonumber 
&\delta\psi_0^j (\varphi;t) =  \h e^{(\omegai_{j}-\frac{i\alpha }{2})t}\sum_{s=0}^{\infty} 
\Big\{e^{i (s \varphi-\omegar t)}\\
&- \left(\frac{a+(-1)^j \sqrt{a^2-b^2}}{b}\right)^* e^{-i [(s-2k_+)\varphi-\omegar t]}\Big\},
\end{align}
and $l=5,6$. 
If $\omegai_5=-\omegai_6>0$, so that $a^2<b^2$, we find that  
\begin{align}
\nonumber
&\delta\psi_0^j (\varphi;t)
= \h e^{\omegai_j t +i(k_+\varphi+\frac{\thetaSpin}{2}-\frac{\alpha t}{2})}\\
&\times \sin\left[(s-k_{+})\varphi - \omegar_j t+\frac{\thetaSpin}{2}\right],
\label{deltarho0}
\end{align}
where 
\begin{align}
\nonumber 
\label{eq_thetaSpin}
\thetaSpin &=-\text{sign}(a)\arctan\sqrt{\left(\frac{b}{a}\right)^2-1}\\ 
&+\frac{\pi}{2}[1-\text{sign}(ab)].
\end{align}

\end{document}